\documentstyle[prl,aps,floats,epsf,epsfig]{revtex}

\begin{document}

\newcommand{\beqa}{\begin{eqnarray}}
\newcommand{\eeqa}{\end{eqnarray}}

\newcommand{\bS}{{\bf S}}
\newcommand{\bB}{{\bf B}}
\newcommand{\bk}{{\bf k}}
\newcommand{\bv}{{\bf v}}
\newcommand{\br}{{\bf r}}
\newcommand{\bp}{{\bf p}}
\newcommand{\s}{{\sigma}}
\newcommand{\bA}{{\bf A}}
\newcommand{\bsig}{\mbox{\boldmath{$\sigma$}}}

\draft

\title{Noise spectroscopy of a single spin with spin polarized STM}
\author{Z. Nussinov$^{1}$, M. F. Crommie$^{2,3}$, and A. V. Balatsky$^{1}$}
\address{$^{1}$ Theoretical Division, Los Alamos National Laboratory, Los
Alamos, NM 87545}
\address{$^{2}$ Department of Physics,
University of California, Berkeley, CA 94720}
\address{$^{3}$ 
Materials Sciences Division, Lawrence Berkeley
Laboratory, Berkeley, CA 94720}

\date{\today}

\twocolumn[

\widetext
\begin{@twocolumnfalse}

\maketitle

\begin{abstract}

We show how the {\em noise} in a spin polarized STM tunneling
current gives valuable spectroscopic information on the temporal
susceptibility of a single magnetic atom residing on a
non-magnetic surface.
\end{abstract}


\vspace{0.5cm}

\narrowtext

\end{@twocolumnfalse}
]

\section{Introduction}

No fundamental principle precludes the measurement of a single
spin, and therefore the capability to make such a measurement
depends on our ability to develop a detection method of sufficient
spatial and temporal resolution. The standard electron spin
detection technique- electron spin resonance- is limited to a
macroscopic number ($\ge 10^{10}$) of electron spins \cite{ESR}.
Recent experiments employing spin polarized STM
\cite{Wiesendanger} open new possibilities of investigation of
magnetic systems at spatial resolutions of the Angstrom scale. The
bulk of the experiments performed to date have been on
magnetically ordered states, such as antiferromagnets.

\begin{figure}[htbp]
\begin{center}
\includegraphics[width = 2.0 in]{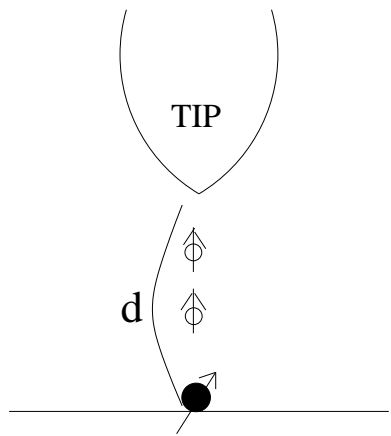} \vspace{0.3cm}
\caption{The experimental setup. The surface tip separation
distance is $d$. The single magnetic atom on the surface is marked
by a solid circle. The spin polarized current emanating from the
STM tip is schematically shown by the spin up open symbols.}
\label{fig:tip}
\end{center}
\end{figure}

Here we propose to use a spin polarized STM tunnel current to gain
spectroscopic information on a single magnetic atom on a
non-magnetic surface. The set up is similar to the one used in
recent ESR-STM experiments, although we now consider the same for
spin polarized current. A scheme is depicted in
Fig.(\ref{fig:tip}). We start with a magnetic atom of spin
$\bf{S}$ placed on an otherwise non-magnetic substrate. As we will
explain in the text, the noise in the current flowing from the STM
tip will allow us, in certain instances, to directly measure the
single spin time dependent susceptibility. This is yet another
case where the spectroscopy of current noise allows us (where
other methods often fail) to directly probe the highly disordered
quantum states of a microscopic system (in this case a single
spin).

In earlier works, one of us and others examined similar issues at
finite magnetic fields. This work is crucially different from its
predecessors in that we consider the situation at {\em a vanishing
external field}, and within the context of a spin polarized tunnel
current. Our work also differs in the 
former context from 
\cite{Bulaevskii}, and
further offers a transparant and accesible
account of the underlying physics. Throughout
the paper we propose and analyze
aspects of a new experimental technique 
for probing single spin dynamics.

The outline of this article is as follows: In section(\ref{main}),
we derive the general relation between the noise $\langle |\delta
I(\omega)|^{2} \rangle$ in the current and the single spin
susceptibility $\chi(\omega)$. Under quite general circumstances,
the noise in the current is directly proportional to the single
spin susceptibility, so that measuring the noise in the current
immediately gives us valuable information about the single spin
susceptibility. In Section(\ref{SPIN}), we will elaborate on the
origin of the spin dependent tunneling matrix elements that form
much of the backbone of our proposal. In Section(\ref{BACK}), we
will discuss the decoherence resulting from backaction effects.
Here, we will also relate the spin scattering relaxation rate and
DC charge transport and estimate how large the various quantities
might be in realistic setups. We will conclude, in
Section(\ref{S/N}), by evaluating the typical signal to noise
ratios for our experimental proposal.

\section{How to probe the Spin Susceptibility via a measurement of
the noise} \label{main}

There is a major difference between probing a single spin (as
discussed here) and probing a macroscopic magnetic system. In the
case of a single atomic spin ${\bf S}(t)$, we perform a
measurement on a microscopic state that is quantum and highly
fluctuating due to interactions with its environment. The
tunneling experiment discussed here is a specific example of {\em
noise spectroscopy}, wherein we extract spectroscopic information,
such as the single spin relaxation time $T^*$ from the measurement
of the noise in the current. The main idea of noise spectroscopy
in STM is that the time dependence of the tunneling current will
have a characteristic relaxation time that is directly related to
spin relaxation time $T^*$. 
Here $T^{*}$ does not originate
from a directly applied external
magnetic field (there is none) 
but rather from substrate excitations
and other effects.

The time varying tunneling current
$I(t)$ will have a fluctuating component, apart from its average
DC value $I_0$.  Current noise is given by the Fourier transform
of the time-dependent fluctuations of the electrical current
leading to the power spectra $\langle |\hat{I}(\omega)|^{2}
\rangle$ within the frequency domain. Here, we will focus on the
power spectrum of the tunneling current and argue that it is
proportional to the local spin susceptibility,
\begin{equation}\label{EQ:suceptibility}
\chi(\omega) = \langle{\bf S}_{\omega} \cdot {\bf
S}_{-\omega}\rangle = \frac{1/T^*}{\omega^2 + 1/T^{*2}},
\end{equation}
of the single atom on the substrate. In the above, $T^{*}$ denotes
the relaxation time to the polarization axis of the single local
spin (for a macroscopic collection of spins, this would be none
other than the standard $T_{1}$ of NMR gauging the relaxation time
to the polarization axis). We reiterate that in the present
discussion  there is a direct externally applied magnetic
field $T^{*}$, and albeit the 
similarity the origin of
the relaxation time here 
and in NMR of a completely
different origin.

A long time exponential relaxation for
$\chi(t) \equiv \langle S(t) S(0) \rangle \sim \exp[-t/T^{*}]$,
has as its  Fourier transform the Lorentzian of
Eqn.(\ref{EQ:suceptibility}). In the Lorentzian form of
Eqn.(\ref{EQ:suceptibility}), the inverse relaxation time $(T^{*})^{-1}$
trivially appears as the linewidth in $\chi(\omega)$. If
additional short time spin modulations occur, these will augment
$\chi(\omega)$ by additional high frequency terms.

In what follows, most of our focus will be on the long time (short
frequency) behavior of the current noise. In thermal noise, the
spectral density $\langle |I(\omega)|^{2} \rangle$ is independent
of frequency (e.g. being $4k_{B}T/R$ at a temperature $T$ for
current flowing through a resistor of magnitude $R$). Shot noise,
on the other hand, results from the discrete pulse-like character
of electrical current. Its magnitude exhibits a white noise
(frequency independent) spectrum that is proportional to the
average DC current, $\langle |\hat{I}(\omega)|^{2} \rangle \simeq
aeI_{0}$ (where the numerical constant $a=2$ for a simple
conductor), up to a certain cut-off frequency which is related to
the time required for an electron to travel through the conductor.

Examples of noise spectroscopy include the noise NQR measurements
\cite{Clarke}, noise in Faraday rotation \cite{Faraday} and,
recently, noise spectroscopy of a local spin dynamics in STM
\cite{ESR-STM}. The central feature present in all of these
examples is that the quantum system is not driven by external
fields. Rather, the noise in the signal itself  (e.g. thermal,
shot, ...) sans any applied polarizing field allows us to extract
spectroscopic information. Here we consider the excess noise produced
by a single spin whose time-dependent quantum state we wish to
probe. We suggest that over a certain parameter range, the excess
noise generated by the single spin will be the dominant noise
source in a single spin tunnel junction.

To make matters concrete, consider the tunneling between two
contacts in the presence of a localized spin $\bS$.  The
Hamiltonian of this system assumes the form
\begin{eqnarray}\label{EQ:Ham}
\hat{H} = \{ \sum_{n\sigma} \epsilon_{L n \sigma} c^\dagger_{L n
\sigma} c_{L n \sigma}  \nonumber
\\ + \sum_{n n^{\prime} \alpha
\beta} c^\dagger_{L n \alpha}[t_0 + t_1 \hat{\bS} \cdot
\hat{\bsig}_{\alpha \beta}]c_{R n^{\prime} \beta}
 \}+
(L \rightarrow R).
\end{eqnarray}
In the above, $\hat{\bsig}_{\alpha \beta}$ is the Pauli matrix
vector with matrix indices $\alpha$ and $\beta$, the fermionic
$c_{\lambda n \sigma},c^{\dagger}_{\lambda n \sigma}$ are the
annihilation and creation operators of electrons in the n-th
eigenstate of the lead $\lambda = L,R$ with $\sigma = \pm 1$ the
(up/down) spin polarization label. The left lead (L) is the STM
tip, and the right lead (R) refers to the surface. For infinite
``free'' leads, the eigenstates $\{|n\rangle\}$ simply correspond
to the various plane wave states $\{|{\bf{k}} \rangle\}$.
 In real
systems, all hopping matrices $\hat{t}$ will carry additional
$n,n^{\prime}$ labels. However, to make the notation more compact
we will dispense with these indices. The wavefunctions of our
system are superpositions of the direct product states
\begin{eqnarray}
|\psi_{L} \rangle \otimes |\psi_{S} \rangle \otimes |\psi_{R}
\rangle
\end{eqnarray}
- the direct product of the state of the left contact, the
impurity spin, and the right contact. The tunneling matrix
$\hat{t}$ present in the second term of Eqn.(\ref{EQ:Ham}) couples
all of these different states. It has two contributions: the term
proportional to $t_0$ describes the spin independent tunneling
while the term proportional to $t_1$ depicts the spin dependent
contributions arising from the exchange interaction for electrons
tunneling to the magnetic atom. Only the second term in
Eqn.(\ref{EQ:Ham}) will give rise to net current flow from the
left to the right contact. In section(\ref{SPIN}), we will explain
the origin and elaborate on the magnitude of the spin independent
and dependent terms. A chemical potential shift (voltage drop)
between the two bands $\{\epsilon_{nL}\}$ and $\{\epsilon_{nR}\}$
will lead to a DC current within the steady state.

Henceforth, we will assume that the tunneling electrons are
partially spin polarized. There are several situations where such
interactions may materialize. In a ferromagnetically coated tip a
chemical potential difference $2(\delta \mu_{\sigma})$ separates
the two different spin polarizations: $\epsilon_{L n \sigma} =
\epsilon_{L n} + \sigma \delta \mu_{\sigma}$. Ferromagnetically
ordered tips have proven to be very successful in the study of
magnetic structures \cite{Wiesendanger}.

Very recently, an alternative approach using 
antiferromagnetically coated tips with
no ferromagnetic order has been used to produce spin polarized
current \cite{AFtip}. These tips might have potential
benefits as compared to ferromagnetic tips. A ferromagnetic tip
may produce a field of ${\cal{O}}(1)$ Tesla at a separation of few angstroms
from the surface. Ferromagnetic tips may therefore lead huge
precession frequencies that are difficult to measure. In the case
of an antiferromagnetic tip, there is a vanishing 
dipolar field.

Both of these techniques may be used for the measurements 
proposed here. In what follows, we are not interested
in a specific model for how the spin polarized current is
generated. We define a parameter $A$ that relates the spin
polarized current to the net tunneling current. It is this parameter 
that will be determined by a particular microscopic model of the tip.
Hereafter, we will treat $A$ as a phenomenological parameter.

As may be seen by examining the spin dependent contribution to
$\hat{t}$, the tunneling electrons exert torques on the localized
spin $\bS$ which lead to corrections to the spin dynamics
\cite{ralph}. To lowest order in $t_{1}$, however, such effects
are not present. Similarly, in what follows,
we ignore the spin-flip interactions between the local moment and 
electrons in the substrate (i.e. we
assume that the experimental temperature is higher than any
relevant Kondo temperature ($T> T_{K}$)).
Below the Kondo temperature, we
may not ignore the spin-flip interactions of the localized spin
with the substrate electrons- the local spin susceptibility will be heavily
influenced by these interactions. Here
we will only consider $T> T_{K}$ (a free 
impurity spin) for the sake of 
simplicity. For temperatures
lower than
the Kondo scale, the Kondo
effect might manifest itself
through interesting changes
in the observed current 
noise that we 
discuss here 
for the free 
spin case.

To make our expressions slightly more concise, we will employ the
Heisenberg representation and absorb the time dependence in all
operators $\{\hat{O}\}$, i.e. $\hat{O}(t) = \exp[iHt] \hat{O}_{S}
\exp[-iHt]$, with $\hat{O}_{S} = \hat{O}(t=0)$ the operator in the
Schroedinger representation which we now forsake. All finite
temperature expectation values $\langle \hat{O}(t) \rangle$ that
appear will represent $\sum_{i} p_{i} \langle \psi_{i}(0) |
\hat{O}(t)| \psi_{i} (0) \rangle$ with $\psi_{i}(t=0)$ the zero
time wavefunction of the Schroedinger representation which does
not evolve within the Heisenberg formulation and $p_{i}$ its
probability within the density matrix formulation.

Let us now give a simple qualitative description of the effect
that we address here. By directly computing $dN_{L}/dt$ we find
that the charge current \cite{explain-current}
\begin{eqnarray}
\hat{I}(t) = - i e \sum_{n n^{\prime} \alpha \beta} c_{L n
\alpha}^{\dagger}(t) [t_0 \delta_{\alpha \beta} + t_1 \hat{\bS}(t)
\cdot \hat{\bsig}_{\alpha \beta}] c_{R n^{\prime} \beta}(t)
\nonumber
\\ + h.c.,
\label{FULL-I}
\end{eqnarray}
with $e$ the electronic charge.

Thus, we see that the tunneling current has a part that depends on
the localized spin via a scalar product,
\begin{equation}\label{eq:I1}
  \delta \hat{I}(t) = e t_1\hat{{\bf S}}(t)  \cdot \hat{{\bf I}}_{spin} (t),
\end{equation}
where
\begin{eqnarray}
\hat{{\bf I}}_{spin} (t) = -i \sum_{ n n^{\prime} \alpha \beta}
c^{\dagger}_{L n \alpha}(t)
 \hat{\bsig}_{\alpha \beta}c_{R n^{\prime}
\beta}(t) + h.c., \label{I-SPIN}
\end{eqnarray}
is the spin polarization dependent contribution to the electronic
current.  This expression has a very transparent meaning. Its
z-component, $\hat{I}^{z}_{spin}(t) = - i (c^{\dagger}_{L,
\uparrow} c_{R, \uparrow} - c^{\dagger}_{L, \downarrow} c_{R,
\downarrow}) + h.c.$, is the net flow of up spin minus the flow of
down spin. A DC current of polarized electrons flowing from the
tip to the surface trivially leads to a uniform spin polarized
current $\hat{{\bf{I}}}_{spin}$. We assume that there is a
non-vanishing {\em steady} spin polarized current component
tunneling from the tip to the surface, assumed to be aligned along
(or defining) the $z$ axis: $\langle  \hat{I}^i_{spin} (t)\rangle
= \delta_{i,z} A^{\frac{1}{2}}$ + time dependent fluctuations,
with a finite $A \neq 0$. An application of the 
diagrammatic analysis introduced in \cite{shnirman},
reveals that four contributions result each of
which is, at most, of the order of 
the contribution that we discuss
below. To lowest order in the hopping amplitude
$t_{1}$, the electronic current-current correlation function
originating from the spin dependent part that we wish to probe,
\begin{eqnarray}\label{EQ:II1}
\langle \{\delta \hat{I}(t),  \delta \hat{I} (t')\}\rangle  =
e^{2} t^2_1 \langle \hat{S}^i(t) \hat{S}^j(t')\rangle \langle
\hat{I}^i_{spin}(t)\hat{I}^j_{spin}(t') \rangle \nonumber
\\ + (t \leftrightarrow t'),
\end{eqnarray}
where $\{ \}$ denotes a symmetrized correlator, $i,j = x,y,z$
denote the spin components. To lowest non-trivial order in $t_1$,
we need to treat the two temporal correlation functions
$\chi(t-t') = \langle \hat{S}^z(t) \hat{S}^z(t')\rangle$ and
$C(t-t') = \langle \hat{I}^i_{spin}(t) \hat{I}^j_{spin}(t')
\rangle \to \langle \hat{I}^i_{spin}(t) \rangle \langle
\hat{I}^j_{spin}(t') \rangle = \delta_{i,z} \delta_{j,z} A$ (for $
{|t-t'|\rightarrow \infty}$) independently. To
this order, the wavefunctions with respect to which we compute the
expectation values are the products of pieces describing the
decoupled evolution of both the spin and of the left and right
contacts separately. To make connection with the main proposal in
this paper, we note that, when Fourier transformed, the
symmetrized correlator $\langle \{\delta \hat{I}(t), \delta
\hat{I}(t')\}\rangle$ is none other than the current noise
spectrum at various frequencies originating from the local spin.
For small $(t_{1}/t_{0}) \ll 1$ (the experimental situation), $
{\cal{O}}(\langle \{\delta \hat{I}(t),  \delta \hat{I}
(t')\}\rangle) = (t_{1}/t_{0})^{2} I_{0}^{2}$. In evaluating the
spin current correlator $C(t)$, we ignore the fluctuating
contributions present for short times (large frequencies). The
spin current correlator $C(t)$ has a finite, asymptotic, long time
value, $A$, that reflects the spin polarized DC current emanating
from the STM tip. In Fourier space, the current
power spectrum is given by a convolution of the two power spectra
associated with $\bS$  (i.e. $\chi(\omega)$) and $\bsig$ (the spin
current correlator $C(\omega)$):
\begin{equation}\label{EQ:II1}
\langle |\delta \hat{I}(\omega)|^{2} \rangle  = e^{2} t^2_1 \int
\frac{d\omega_1}{2\pi} \chi(\omega_1) C(\omega - \omega_1) +
(\omega \rightarrow -\omega).
\end{equation}
At low frequencies,
 $C(\omega) \simeq 2\pi A \delta(\omega)$, and, consequently,
\begin{equation}\label{EQ:II2}
\langle |\delta \hat{I}(\omega)|^{2} \rangle  = 2 A  e^{2} t^2_1
\chi(\omega) + ...
\end{equation}

The ellipsis in Eqn.(\ref{EQ:II2}) refer to the contribution to
the convolution of Eqn.(\ref{EQ:II1}) from the finite frequency
(short time) contributions to $C(t)$. Assuming such finite frequency
contributions in $C(\omega)$ have low spectral weight, the
effect of these contributions will be low. This low order result
is augmented by higher order corrections in $t_{1}$ as well as
contributions arising from the connected component of the current
current correlator which amounts to a shot noise like
contribution.

As noted, in Eqn.(\ref{EQ:II2}) we neglected
the effect of finite frequency
components of $C(\omega)$ in the convolution
of Eqn.(\ref{EQ:II1}).
We believe the large finite
frequency
components of the spin current 
$C(\omega)$ to be small as these
correspond to transient fluctuations about an assumed steady state.  
This assumption is not necessary, however. 
More generally, we may deal with the full
convolution in Eqn.(\ref{EQ:II1}) directly without invoking any assumptions. 
We may potentially do this by
first performing a measurement
on a reference state.
For instance, if we initially replace the 
single spin by a large cluster of a large fixed
spin, we may then experimentally measure the resulting
spin current $C(\omega)$.
Armed with the
knowledge of the spin
current
(from this earlier measurement
on the magnetic cluster), 
once the noise spectrum 
$\langle |I(\omega)|^{2} \rangle$ is
measured for the single spin, we can directly 
deconvolve Eqn.(\ref{EQ:II1}) to
obtain the spin susceptibility \cite{C_w}.

Eqs.(\ref{EQ:II1}, \ref{EQ:II2}) are our central result. 
They vividly illustrate
how the spectroscopy of the {\em noise} in the tunneling current
$\langle |\delta I(\omega)|^{2} \rangle$ allows us to directly
probe the spectrum of spin fluctuations encapsulated in
$\chi(\omega)$. The spin polarized tunneling current provides a
reference frame with respect to which we may measure the
fluctuations of the localized spin $\bS(t)$. We note that these
results are generally applicable to systems displaying strong
correlations (such as impurities in the Kondo regime).

\section{The origin of Spin Dependent Tunneling}
\label{SPIN}

We now elaborate on the origin and magnitude of the spin dependent
tunneling matrix elements of Eqn.(\ref{EQ:Ham}). The spin
dependence of the tunneling originates from the direct exchange
dependence of the tunneling barrier \cite{ESR-STM}. The overlap of
the electronic wave functions of the tip and surface, separated by
a distance $d$ is exponentially small and is given by a {\em spin
dependent} tunneling matrix element,
\begin{eqnarray}\label{eq:G}
  \hat{t} = \gamma \exp[-\sqrt{\frac{\Phi -
J{\bf S}(t) \cdot
  {\hat{\bsig}}}{\Phi_0}}],
\end{eqnarray}
where we explicitly include the direct exchange between tunneling
electron spin $\bsig$ and the local spin $\bS$. Here, $J$ is the
exchange interaction between the electrons tunneling from the tip
and the local precessing spin $\bS$. In the above, $\hat{t}$ is to
be understood as a matrix in the internal spin indices, and $\Phi$
is the tunneling barrier height. Typically, $\Phi$ is a few eV. As
a canonical value we may assume $\Phi = 4 eV$, $\Phi_0 =
\frac{\hbar^2}{8 m d^2}$ is related to the distance $d$ between
the tip and the surface \cite{Stroscio}. As the exchange term in
the exponent is small compared to the barrier height, we may
expand the exponent in $JS$. Explicitly, $\hat{t}$ may be written
as
\begin{eqnarray}\label{eq:t2}
\hat{t} = t_0 + t_1 \hat{\bsig}  \cdot {\bf S(t)},
\end{eqnarray}
where,
\begin{eqnarray}
t_0 = \gamma \exp(-(\Phi/\Phi_0)^{1/2}) \cosh[\frac{JS}{2\Phi}
\sqrt{\frac{\Phi}{\Phi_0}}],
\end{eqnarray}
describes spin independent tunneling. The spin dependent
amplitude,
\begin{eqnarray}
t_1 = \gamma \exp(-(\Phi/\Phi_0)^{1/2}) \sinh[\frac{JS}{2\Phi}
\sqrt{\frac{\Phi}{\Phi_0}}].
\end{eqnarray}
For estimates we may employ the typical rule of thumb $t_1/t_0
\simeq \frac{JS}{2\Phi} \ll 1$.

\section{Backaction effect of the tunneling current on the spin}
\label{BACK}

We may use the tunneling Hamiltonian of Eqn.(\ref{EQ:Ham}) to
estimate the decay rate of the localized spin state resulting from
the spin scattering interaction associated with $t_1$. To second
order this calculation is equivalent to a simple application
Fermi's golden rule leading to an up-down spin flip rate
$\frac{1}{\tau_s} = \pi t_1^2 N_L N_R eV$, with $V$ is the voltage
applied between the left and right electrodes. Similarly, the DC
tunneling current $I_0$ is given by the tunneling rate of
conduction electrons $ \frac{1}{\tau_e} = \pi t_0^2 N_L N_R eV$,
where $N_{L,R}$ denotes the density of states at the Fermi level
of the tip and surface respectively \cite{Korotkov}.

Diagrammatically, both the spin scattering and electronic
scattering rates arise from the simple bubble diagram whose
real-time propagators are $G_{L}(t)$ and $G_{R}(-t)$, where
$G_{L,R}$ correspond to the left and right Green's functions for
the conduction electrons respectively. The sole difference between
the two (spin dependent/independent) scattering rates is
encapsulated in the prefactors. In the spin dependent scattering
case ($\tau_{s}^{-1}$) the raw value of the single loop integral
appearing in the bubble diagram needs to be scaled by $t_{1}^{2}$.
The spin independent scattering rate ($\tau_{e}^{-1}$) is the much
same albeit a scaling by the spin independent scattering amplitude
squared ($t_{0}^{2}$).

Comparing these simple results, we find
\begin{eqnarray}
\frac{1}{\tau_{s}} = (\frac{t_{1}}{t_{0}})^{2} \frac{1}{\tau_{e}}
\simeq (\frac{JS}{2 \Phi})^{2} \frac{1}{\tau_{e}}.
\label{linewidth}
\end{eqnarray}

The important outcome of this analysis is that the current induced
broadening predicts a spin relaxation rate $\frac{1}{\tau_s}
\propto I_0$ which may be experimentally tested. This result has a
very simple interpretation: the impinging foreign electron
tunneling rate for a DC current of magnitude $I_0 = 1 nA$ is given
by $\frac{1}{\tau_e} \sim 10^{10} Hz$. By contrast, the
probability to produce a spin flip, sparked by the tunneling
electrons, is proportional to $t_1^2$, which leads to
Eqn.(\ref{linewidth}) for the linewidth. The full intrinsic line
width is further enhanced by the coupling of the spin to the
environment (e.g. the interaction between the spin and various
substrate excitations) which may indeed further increase the spin
flip rate. The net observed linewidth,
\begin{eqnarray}
(T^{*})^{-1} \simeq \tau_{s}^{-1} + \tau_{env}^{-1},
\label{env+spin}
\end{eqnarray}
includes both backaction contributions ($\tau_{s}^{-1}$) and the
aforementioned linewidth broadening due to coupling to the
environment ($\tau_{env}^{-1}$). The inverse backaction relaxation time
scale sets a lower bound on the net relaxational linewidth of the
single impurity spin.

Given the typical values of the parameters in
Eqn.(\ref{linewidth}), we estimate $\frac{1}{\tau_s} \sim 5 \times
10^{6} Hz $ for our DC current of $I_{0} \sim 1 nA$, $J \sim 1$
eV, $\Phi \sim 4$ eV, and $S = 1/2$. Future experiments will help
to clarify the linewidth dependence on the various parameters.

\section{A Sizable Signal to Noise Ratio}
\label{S/N}

We now demonstrate that the (signal to noise) ratio of 
$|\delta(\omega \to 0)|^{2}$ (the excess noise induced by
the impurity spin) to $|I_{shot} (\omega \to 0)|^{2}$ 
(the shot noise already present in the
absence of the impurity spin) can be of quite significant
(of order unity). 

The finite frequency ratio $|\delta I(\omega)|^{2}/| I_{shot}(\omega)|^{2}$ 
was computed in a multitude of
systems and shown to be bounded by 4 or other
numbers of order unity. By contrast, the low or zero frequency 
signal to noise ratio $|\delta I(\omega \to 0)|^{2}/
| I_{shot}(\omega \to 0)|^{2}$ was found to be unbounded
in many instances. For one calculation amogst many others
demonstrating this explicitly, see e.g. \cite{makhlin}. 
Nevertheless, as we highlight below, the low frequency signal to noise
ratio,  $|\delta I(\omega \to 0)|^{2}/
| I_{shot}(\omega \to 0)|^{2}$, in our system is
stringently bounded from 
above by a number of order unity,
just as it is in many 
finite frequency situations. 
By considering typical parameter
values, we will show that this ratio may, potentially, 
saturate this upper bound and be markedly large. The spin dynamics as 
manifested through the current noise may be very 
easily discernible.

To estimate this ratio, we note that the average spin dependent
contribution to the current,
\begin{eqnarray}
\langle \delta \hat{I} (t) \rangle = e t_{1} \langle S^{i}(t)
\hat{I}^{i}_{spin}(t) \rangle
\end{eqnarray}
leading to ${\cal{O}}(e t_{1} A^{1/2} \langle S^{z}(t)
\rangle)$. Here, as throughout, $\hbar =1$, 
and the spin is dimensionless. 

To obtain estimates of orders of magnitude let us inspect
Eqs.(\ref{FULL-I}, \ref{I-SPIN}). The net electronic current
$I_{e}$ of Eqn.(\ref{FULL-I}) is of the order of $I_{0}$. Insofar
as orders of magnitude are concerned, the spin current
($I_{spin}$) defined in Eqn.(\ref{I-SPIN}) satisfies
\begin{eqnarray}
{\cal{O}}(A^{1/2}) = {\cal{O}}(I_{spin}) =
{\cal{O}}(\frac{I_{e}}{et_{0}}) =
{\cal{O}}(\frac{I_{0}}{et_{0}}).
\end{eqnarray}

The shot noise, $\langle I_{shot}^{2} \rangle = aeI_{0}$, where
the numerical constant $a= {\cal{O}}(1)$. Noting that $I_{0} =
\frac{e}{\tau_{e}}$, and making use of  Eqn.(\ref{EQ:II2}), the
signal to noise ratio is found to be
\begin{eqnarray}
\frac{\langle |\delta \hat{I}(\omega)|^{2}\rangle}{\langle |\delta
I_{shot}(\omega \to 0)|^{2} \rangle} \sim
\frac{t_{1}^{2}}{t_{0}^{2}} \chi(\omega) \frac{1}{\tau_{e}}.
\end{eqnarray}
The ratio $(t_{1}/t_{0}) \simeq \frac{J}{2\Phi}$. Inserting the
Lorentzian form of $\chi(\omega)$ from
Eqn.(\ref{EQ:suceptibility}), we obtain
\begin{eqnarray}
\frac{\langle |\delta I (\omega \to 0)|^{2}\rangle}{\langle
|\delta I_{shot} (\omega \to 0)|^{2} \rangle} \sim
\frac{T^{*}}{\tau_{e}} (\frac{t_{1}}{t_{0}})^{2}\nonumber
\\ \sim (\frac{J}{2\Phi})^{2} \frac{T^{*}}{\tau_{e}}.
\label{loose}
\end{eqnarray}

Inserting typical values, $J \simeq 0.1 -1$ eV, $\Phi \simeq 4$ eV,
$T^{*} \sim 10^{-8} - 10^{-6}$ seconds, and $\tau_{e} \sim
10^{-10}$ seconds (1nA), we find the above
signal to noise ratio is, naively, 1-100.  
As promised, we now demonstrate, within
this allowed  empirical regime, 
the signal to noise ratio
will typically veer
towards the lower
end of the spectrum
(i.e. may be of order unity
at most). The bottleneck in the
signal to noise
ratio is set
by the backaction
effect of the single
spin on the tunneling current.
More explicitly, fusing 
Eqs.(\ref{linewidth},\ref{env+spin})
together,
\begin{eqnarray} 
(T^{*})^{-1} \simeq \tau_{s}^{-1} + \tau_{env}^{-1} 
\nonumber 
\\ 
\simeq
(\frac{JS}{2 \Phi})^{2} \frac{1}{\tau_{e}}
+ \tau_{env}^{-1},
\end{eqnarray}
leading us to conclude 
that $T^{*} \lesssim  (\frac{2 \Phi}{JS})^{2} \tau_{e}$.

Inserting this back in the last 
line of Eqn.(\ref{loose}),
\begin{eqnarray}
\frac{\langle |\delta I (\omega \to 0)|^{2}\rangle}{\langle
|\delta I_{shot} (\omega \to 0)|^{2} \rangle}
\lesssim (\frac{J}{2\Phi})^{2} \frac{T^{*}}{\tau_{e}} 
\lesssim {\cal{O}}(1).
\end{eqnarray}

As the generic signal to noise ratio
is bounded both from below and
above by a number of order
unity, the signal
may indeed be quite
sizable (of the order
of the shot noise)
and may be 
detected.

This sizable ratio offers promise for such
single spin detection and related small system
applications.

 {\em Note added.} While finishing this paper we became
aware of work by Bulaevskii, Hrushka and Ortiz \cite{Bulaevskii},
where a similar problem was considered for a specific case of
a ferromagnetic STM tip. Our results are qualitatively similar to
the results obtained by Bulaevskii et al.

\section{Acknowledgments}

This work was supported by the US Department of Energy. We are
grateful to Matthew Hastings, Xinghua Lu, Yishai Manassen, A. Shnirman,
and J. X. Zhu, for useful discussions. This work was supported by LDRD
X1WX at Los Alamos (ZN and AVB) and by NSF EIA-0205641 (MFC).



\begin{thebibliography}{99}

\bibitem{ESR}  M. Farle, { Rep. Prog. Phys.} {\bf 61}, 755 (1998).

\bibitem{Wiesendanger} R. Wiesendanger, I. V. Shvets, D. Burgler,
G. Tarrach, H. J. Guntherodt, J. M. D. Coey, S. Graser, Science
{\bf 255}, 583 (1992); M. Bode, M. Getzlaff and R. Wiesendanger
Phys. Rev. Lett. 81, 4256 (1998); S. Heinze, M. Bode, A. Kubetzka,
O. Pietzsch, X. Nie, S. Blugel, R. Wiesendanger, Science {\bf
288}, 1805 (2000).

\bibitem{Bulaevskii} L.N. Bulaevskii, M. Hruska, G. Ortiz, cond-mat/0212049

\bibitem{AFtip}A. Wachowiak, J. Wiebe, M. Bode, O. Pietzsch,
M. Morgenstern, R. Wiesendanger, Science, {\bf 298}, 577, (2002);
A. Kubetzka, M. Bode, O. Pietzsch, and R. Wiesendanger, Phys. Rev.
Lett. {\bf 88}, 057201, (2002).

 

\bibitem{Clarke} T. Sleator, E. L. Hahn, C. Hilbert and J. Clarke, Phys. Rev.
Lett. 55, 1742 (1985); T. Sleator, E. L. Hahn, C. Hilbert and J.
Clarke, Phys. Rev. B 36, 1969 (1987).


\bibitem{Faraday}  E. B.
Aleksandrov and V. S. Zapasskii, Sov. Phys. JETP 54, 64 (1981). T.
Mitsui, Phys. Rev. Lett. 84, 5292 (2000).



\bibitem{ESR-STM}
A.V. Balatsky, Y. Manassen and R. Salem, Phil. Mag. {\bf B 82},
 1291, (2002); A.V. Balatsky, Y. Manassen and R. Salem, Phys.
Rev {\bf B 66}, 195416, (2002).




\bibitem{ralph} D. Ralph, Science, {\bf 291}, 999 (2001),
and references therein.

\bibitem{explain-current} 
The number operator $N_{L} = c_{L}^{\dagger} c_{L}$. 
With $e$ the electronic charge,
The associated charge current 
\begin{eqnarray}
e \frac{dN_{L}}{dt} = -ie[N_{L},H] = -ie [c_{L \sigma}^{\dagger} c_{L \sigma},H]
\nonumber
\\
\to -ie[c_{L \sigma}^{\dagger} c_{L \sigma} ,c_{L}^{\dagger} \hat{t} c_{R} + 
h.c.]
\end{eqnarray}
Taking note of the trivial 
$[c_{L}^{\dagger} c_{L} , c_{L}^{\dagger}] = c_{L \sigma}^{\dagger}$
and $[c^{\dagger} c,c] = -c$,
the charge current
\begin{eqnarray}
e \frac{dN_{L}}{dt} = -ie[c_{L}^{\dagger} \hat{t} c_{R}]
+ h.c.
\end{eqnarray}

 
\bibitem{shnirman}
A. Shnirman, D. Mozyrisky 
and I. Martin, 
cond-mat/0211618

\bibitem{C_w}
The resulting spin current for
the magnetic
cluster will be
different than
that for a single
atom. However, the difference
in the respective spin
currents $C(\omega)$
between the two
cases are sparked 
by backaction
effects (the respective effects
of the cluster and
the single spin on the
tunneling
current) which
are of higher order
in $t_{1}$.

\bibitem{Stroscio} J. Tersoff and N.D. Lang, ''Theory
of Scanning Tunneling Microscopy", in Methods of Experimental
Physics, v 27, 1, Academic Press, 1993.

\bibitem{Korotkov} A. N. Korotkov, D. V. Averin, and
K. K. Likharev, Phys. Rev. {\bf B 49}, 7548 (1994). See also D.
Mozyrsky, L. Fedichkin, S. A. Gurvitz, G. P. Berman,
cond-mat/0201325.

\bibitem{makhlin}
Y. Makhlin, G. Schon, and A. Shnirman,
Phys. Rev. Lett. {\bf{85}}, 4578 (2000)



 

 

\end{thebibliography}
\end{document}